\documentclass[aps,prl,twocolumn,showpacs]{revtex4}
\usepackage[]{graphicx}% eps files
\usepackage{bm}% bold math
\usepackage{amsmath,amssymb,graphicx,bbold}

\begin{document}

\title{Topological phase structure of vector vortex beams}

\author{C. E. R. Souza$^{1}$, J. A. O. Huguenin$^{2}$, and A. Z. Khoury$^{1}$}

\affiliation{$^{1}$
Instituto de F\'\i sica, Universidade Federal Fluminense,
24210-346, Niter\'oi - RJ, Brazil}
\affiliation{$^{2}$
Instituto de Ci\^encias Exatas, Universidade Federal Fluminense, 
27213-415, Volta Redonda - RJ, Brazil}

%\date{\today}
 
\begin{abstract}
The topological phase acquired by vector vortex optical beams is investigated. 
Under local unitary operations on their polarization 
and transverse degrees of freedom, the vector vortices can only acquire discrete 
geometric phase values, $0$ or $\pi\,$, associated with closed paths belonging to 
different homotopy classes on the SO($3$) manifold. These discrete values are 
demonstrated through interferometric measurements and the spin-orbit mode 
separability is associated to the visibility of the interference patterns. 
The local unitary operations performed on the vector vortices involved both 
polarization and transverse mode transformations with birefringent wave plates 
and astigmatic mode converters. The experimental results agree with our 
theoretical simulations and generalize our previous results obtained with 
polarization transformations only. 
\end{abstract}
\vskip2pc 
 
\maketitle

\section{introduction}

Since the seminal works by Pancharatnam \cite{pancha1,pancha2}, Aharonov \cite{aharonov} and 
Berry \cite{berry}, the concepts of geometric and topological phases found numerous applications 
in different contexts. Besides its intrinsic beauty, geometric phases are candidates for robust 
implementations of quantum computing gates \cite{vedral,zoller}. 
The geometric phase acquired by optical vortices under cyclic transformations was theoretically 
predicted by van Enk \cite{vanenk} and interpreted in terms of an orbital Poincar\'e sphere 
proposed by Padgett and Courtial \cite{padgett}. Later, this phase was experimentally 
demonstrated by Galvez and co-workers through interferometric measurements \cite{galvez}.  
The geometric phase on entangled bipartite systems was extensively discussed in 
Refs. \cite{erik1,erik2,erik3,erik4}.  
The topological structure of the geometric phase acquired by entangled quantum states 
under cyclic evolutions was investigated by P. Milman and R. Mosseri 
\cite{remy,milman}, where they studied the role played by the 
topology of the SO($3$) group in connection with maximally entangled states (MES). 
The elements of the SO($3$) group can be represented on a parameter space with 
nontrivial topology, a sphere of radius $\pi$ and its diametrically opposite points 
identified. Two homotopy classes of closed trajectories can be identified in this 
topology. One is composed by closed trajectories that cross the 
surface of the sphere an even number of times, which we shall designate as 
the $0-$class ones. Closed trajectories that cross the 
surface of the sphere an odd number of times will be called the $\pi-$class ones. 
Each point of the SO($3$) parameter space can be associated to a maximally 
entangled state as in ref.\cite{remy2}. The relevance of this parameterization is 
evidenced when the geometric phase is computed for cyclic transformations performed 
on maximally entangled states. 
The topological phase acquired by a maximally entangled state is $0$ for $0-$class 
and $\pi$ for $\pi-$class trajectories. This effect was observed almost simultaneously 
in two different systems, spin-orbit transformations on a laser beam \cite{nossoprl} 
and two qubit manipulation by nuclear magnetic resonance (NMR) \cite{nmr}. 
Later, the topological phases were generalized to pairs of qudits of any 
dimension \cite{fracuff} and multiple qubit systems \cite{multiqubit}, 
where fractional phase values were predicted. These fractional topological 
phases can in principle be measured in spatial qubits and qudits encoded 
in quantum correlated photons generated by spontaneous parametric down 
conversion \cite{sebastiao,erik}. 

The study of vector vortex beams as non separable spin-orbit modes, benefit from 
several concepts of quantum entanglement theory. One remarkable example is the 
representation of maximally entangled states in the doubly connected manifold 
associated with the elements of the SO($3$) group, which also holds for vector 
vortices. This representation was used in Ref.\cite{nossoprl} to demonstrate the 
discrete topological phases acquired by a vector vortex beam under polarization 
transformations. 
In the present work we improved our experimental result on spin-orbit laser modes by 
implementing a trajectory where both polarization (spin) and orbital transformations 
are performed. 
Also, we avoided paths lying on the surface of the SO($3$) sphere, where the 
representation is singular. Now the $\pi-$class trajectory implemented crosses the 
surface of the sphere on a single point. Our present results generalize the 
previous ones of Ref.\cite{nossoprl}.

\section{spin-orbit mode structure}

\begin{figure}
\begin{center} 
\includegraphics[scale=.4]{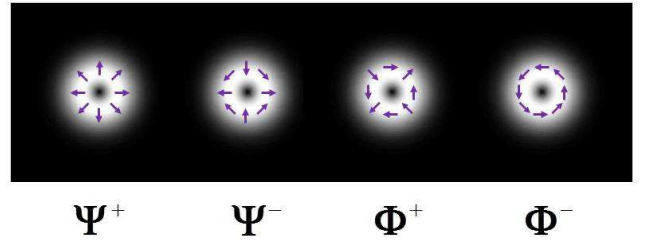}
\end{center} 
\caption{Linear polarization patterns of cylindrically polarized beams.}
\label{cylbeams}
\end{figure}

In many textbooks, the spatial mode structure produced by laser resonators are usually 
described by solutions of the paraxial wave equation \cite{siegman,yariv}. These solutions 
are represented by Hermite-Gaussian (HG) functions in Cartesian coordinates and by 
Laguerre-Gaussian (LG) functions in cylindrical coordinates. 
Both families are characterized by a pair of integer 
indexes and can be cast in a hierarchy of different orders, starting with the zero order 
composed solely by the fundamental Gaussian beam. In one of the seminal works describing 
linear transformations between the two set of modes, Abramochkin and Volostnikov showed how 
the two families of modes are connected. In particular, the first order subspace is 
isomorphous to polarization modes, what allowed Padgett and Courtial to define an 
\textit{orbital} Poincar\'e sphere, where the first order LG modes with $\pm 1$ 
helicity play the role of right and left circularly polarized light, being represented on
the poles of the sphere. The HG modes with all possible orientations 
play the role of linear polarization states and are represented along the equator. 
A first order LG mode propagating along the $z$ direction is described by the 
following wave functions
\begin{eqnarray}
\psi_{\pm}(\rho,\varphi,z)&=&\frac{2}{\sqrt{\pi}}\,\frac{\rho}{w^2(z)}\,
\exp\left(-\frac{\rho^2}{w^2(z)}\right)\\
&\times&\exp\left\{
i\left[\frac{k\rho^2}{2R(z)}+2\arctan\left(\frac{z}{z_R}\right)\right]\right\}\,e^{\pm i\varphi}\;,
\nonumber
\end{eqnarray}
where ($\rho,\varphi$) are the polar coordinates on the transverse plane, $z_R$ is the Rayleigh 
length, $R(z)=(z_R^2+z^2)/z$ is the wave front radius, and 
$w(z)=\sqrt{2\,(z_R^2+z^2)/(k\,z_R)}$ is the beam diameter at position $z\,$. 
In Cartesian 
coordinates, the paraxial wave equation gives rise to the so called Hermite-Gaussian 
(HG) modes. The first order modes are 
\begin{eqnarray}
\psi_h(x,y,z)&=&
\sqrt{\frac{2}{\pi}}\,\frac{2\,x}{w^2(z)}\,
\exp\left(-\frac{x^2+y^2}{w^2(z)}\right)\\
&\times& \exp\left\{
i\left[\frac{k\,(x^2+y^2)}{2R(z)}+2\arctan\left(\frac{z}{z_R}\right)\right]\right\}\;,
\nonumber\\
\psi_v(x,y,z)&=&
\sqrt{\frac{2}{\pi}}\,\frac{2\,y}{w^2(z)}\,
\exp\left(-\frac{x^2+y^2}{w^2(z)}\right)\\
&\times& \exp\left\{
i\left[\frac{k\,(x^2+y^2)}{2R(z)}+2\arctan\left(\frac{z}{z_R}\right)\right]\right\}\;. 
\nonumber
\end{eqnarray}
One easily verifies that the first order LG and HG modes are connected by the 
following simple relation
\begin{eqnarray}
\psi_\pm=\frac{\psi_h\pm i\,\psi_v}{\sqrt{2}}\;. 
\end{eqnarray}
HG modes can also be defined on rotated coordinates systems. For example, the 
HG modes oriented at $\pm 45^\circ$ are given by 
\begin{eqnarray}
\psi_{\pm 45^\circ}=\frac{\psi_h\pm \,\psi_v}{\sqrt{2}}\;. 
\end{eqnarray}
Therefore, if we combine the structure of first order paraxial modes with polarization, 
we can build spin-orbit modes and establish the analogy with two-qubit systems in quantum 
mechanics. For example, cylindrically polarized beams have the same mathematical structure 
of the so called Bell states, defined in connection with Bell inequalities. In terms of 
HG and linear polarization modes they are given by 
\begin{eqnarray}
\nonumber\\
\mathbf{\Psi^\pm}&=&
\frac{\psi_{h}\,\mathbf{\hat{e}}_{_{H}} \pm \psi_{v}\,\mathbf{\hat{e}}_{_{V}}}
{\sqrt{2}}\;,
\nonumber\\
\mathbf{\Phi^\pm}&=&
\frac{\psi_{h}\,\mathbf{\hat{e}}_{_{V}} \pm \psi_{v}\,\mathbf{\hat{e}}_{_{H}}}
{\sqrt{2}}\;.
\label{bell}
\end{eqnarray}
where $\mathbf{\hat{e}}_{_H}$ and $\mathbf{\hat{e}}_{_V}$ are the linear polarization 
unit vectors along the horizontal and vertical directions, respectively.
These modes correspond to the linear polarization patterns depicted in Fig.(\ref{cylbeams}). 
They have been employed in a number of optical experiments both in applied and fundamental 
subjects \cite{leuchs}. Their connection with Bell inequalities both in classical and quantum optics 
has been discussed in Refs.\cite{spreew1,spreew2,belluff,bellchines,bellpadgett,bellsaleh} and 
their use on alignment free quantum cryptography was investigated in 
Refs.\cite{aolita,bb84uff,bb84napoli}. 

%
%============================================================================
%============================================================================
%

For our experimental context, it will be particularly useful to employ the 
LG basis for the spatial modes, since their characteristic phase singularity will 
work as a spatial reference for our interferometric measurements. Therefore, 
it will be convenient to write a general first order spin-orbit mode as:
\begin{eqnarray}
\mathbf{\Psi}&=& 
\lambda_1\,\psi_+\mathbf{\hat{e}}_H +
\lambda_2\,\psi_-\mathbf{\hat{e}}_H +
\lambda_3\,\psi_+\mathbf{\hat{e}}_V 
%\nonumber\\ &+&
+ \lambda_4\,\psi_-\mathbf{\hat{e}}_V
\nonumber\\
&\equiv& \left[\lambda_1,\lambda_2,\lambda_3,\lambda_4\right]^T\;. 
\label{genmode}
\end{eqnarray}
The separability of a spin-orbit mode can be quantified by the analogous 
definition of concurrence. For the spin-orbit mode described by 
Eq.(\ref{genmode}), the concurrence is
\begin{equation}
C=2\mid \lambda_1\,\lambda_4 - \lambda_2\,\lambda_3\mid\;.
\label{conc}
\end{equation}

\section{The experiment}

In order to compute the whole sequence of spin-orbit mode transformations, 
it is essential first to define the operators that describe the 
spatial and polarization mode conversions. We will be dealing with 
wave plates for polarization and astigmatic mode converters for the transverse 
mode transformations, that is, elements acting on each degree of freedom separately. 
We will express rotation angles in degrees and phase retardations in radians for 
immediate identification of their physical meaning. When their fast axis is oriented 
along the horizontal direction, wave plates introduce a retardation phase $\phi$ 
between $H$ and $V$ polarizations. For example, a quarter wave plate (QWP) corresponds to 
$\phi=\pi/2$ and a half wave plate (HWP) to $\phi=\pi$. 
When the fast axis is rotated by the angle $\theta\,$, its action in the 
$\{\mathbf{\hat{e}}_{_H},\mathbf{\hat{e}}_{_V}\}$ basis is described by 
the following SU($2$) matrix 
\begin{eqnarray}
W(\theta,\phi) &=& \left[
\begin{matrix}
\cos\frac{\phi}{2}+i\sin\frac{\phi}{2}\cos 2\theta & i\sin\frac{\phi}{2}\sin 2\theta\\
i\sin\frac{\phi}{2}\sin 2\theta & \cos\frac{\phi}{2}-i\sin\frac{\phi}{2}\cos 2\theta
\end{matrix}
\right]\;.
\nonumber\\
\label{W}
\end{eqnarray}
Spatial mode converters can be made with cylindrical lenses \cite{converter,converter2,converter3,cnot} for variable 
retardation $\phi$, or the DP for $\phi=\pi$. They act on spatial modes $\psi_h$ and $\psi_v$ 
in the same way the wave plates act on horizontal and vertical polarization. However, as stated 
above, we will use the $\{\psi_+,\psi_-\}$ basis for the spatial modes. In this basis a spatial mode converter 
introducing a retardation $\phi\,$, and rotated by the angle $\theta$ is described by the 
SU($2$) matrix 
\begin{eqnarray}
C(\theta,\phi) = \left[
\begin{matrix}
\cos\frac{\phi}{2} & i\sin\frac{\phi}{2}\,e^{-2\,i\,\theta}\\ 
i\sin\frac{\phi}{2}\,e^{2\,i\,\theta} & \cos\frac{\phi}{2} 
\end{matrix}
\right]\;.
\nonumber\\
\label{C}
\end{eqnarray}

\begin{figure}
\begin{center} 
\includegraphics[scale=.35]{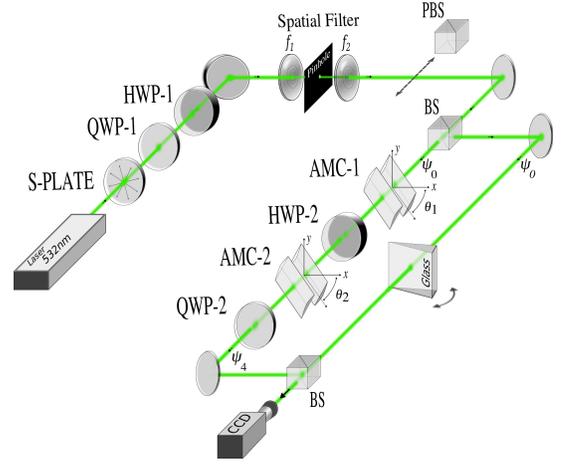}
\end{center} 
\caption{Experimental setup.}
\label{setup}
\end{figure}

The mode preparation is sketched in Fig.\ref{setup}. A linearly polarized 
TEM$_{00}$ mode from a frequency doubled Nd:YAG laser ($\lambda=532\,nm$) 
is transmitted through a radial polarization converter (S-Wave Plate, Model 
ALTECHNA RPC-515-04) that prepares the initial $\mathbf{\Psi^+}$ mode. 
Before sending it through the transformation sequence, the quality of the 
radially polarized beam is checked with a half wave plate and a polarizer beam 
splitter (not shown in the figure). When the beam is well prepared, we observe 
a rotating HG mode transmitted through the polarizing beam splitter, as we 
rotate the wave plate. Then, the test elements are removed from the setup and 
the beam is sent through the transformation sequence. 
The $\mathbf{\Psi^+}$ mode prepared by the S-plate is sent through a 
quarter wave plate (QWP-1) oriented at $0^\circ\,$ and a half wave plate 
(HWP-1) oriented at $22.5^\circ\,$. Then, a spatial filter is used to improve 
the beam shape. This sequence prepares the initial nonseparable mode 
\begin{eqnarray}
\mathbf{\Psi}_0&=&
\frac{\psi_+\,\mathbf{\hat{e}}_{_H}+\psi_-\,\mathbf{\hat{e}}_{_V}}{\sqrt{2}}
\nonumber\\
&\equiv& \frac{1}{\sqrt{2}}\left[1,0,0,1\right]^T\;, 
\label{nsmode}
\end{eqnarray}
for which $C=1\,$. Separable modes were produced by filtering one of the 
polarization components with a polarizing beam splitter (PBS). 
Therefore, the initial mode can be described by the general formula 
\begin{eqnarray}
\mathbf{\Psi}_i&=&
\sqrt{\epsilon}\,\psi_+\,\mathbf{\hat{e}}_{_H}+\sqrt{1-\epsilon}\,\psi_-\,\mathbf{\hat{e}}_{_V}
\nonumber\\
&\equiv& \left[\sqrt{\epsilon},0,0,\sqrt{1-\epsilon}\right]^T\;, 
\label{initialnsmode}
\end{eqnarray}
where $\epsilon=0$ when the horizontal polarization is blocked, $1$ when the vertical 
polarization is blocked and $1/2$ when both polarization are present. 

%
%============================================================================
%============================================================================
%

Let us describe the transformation sequence performed on mode $\mathbf{\Psi}_0\,$, 
that is when $\epsilon=1/2\,$. The probe beam is sent through a Mach-Zehnder (MZ) 
interferometer used for the topological phase measurement. 
In one arm of the interferometer, the initial mode is used as a reference beam. 
In the other arm, a sequence of spatial and polarization mode transformations is 
implemented. First, the initial spin-orbit mode $\Psi_0$ is transmitted through a 
$\pi/2$ astigmatic mode converter (AMC-1) oriented at $22.5^\circ$ that produces mode
\begin{eqnarray}
\mathbf{\Psi}_1 &=& \left[\mathbb{1}\otimes C(22.5^\circ,\pi/2)\right] \mathbf{\Psi}_0 
\nonumber\\
&=& 
\frac{1}{2}\,\left[1,e^{\frac{3i\pi}{4}},e^{\frac{i\pi}{4}},1\right]^T\;.
\label{nsmodepsi1}
\end{eqnarray}
A half wave plates (HWP-2) with a variable orientation $\theta\,$, produces mode 
\begin{eqnarray}
\mathbf{\Psi}_2&=& \left[ W(\theta,\pi)\otimes\mathbb{1}\right] \mathbf{\Psi}_1 
\nonumber\\
&=& 
\frac{1}{2}\left[
\begin{matrix}
i\,\cos2\theta&-&\sin2\theta\,e^{-\frac{i\pi}{4}}\\ 
-\cos2\theta\,e^{\frac{i\pi}{4}}&+&i\,\sin2\theta\\ 
i\,\sin2\theta&+&\cos2\theta\,e^{-\frac{i\pi}{4}}\\ 
-\sin2\theta\,e^{\frac{i\pi}{4}}&-&i\,\cos2\theta
\end{matrix}
\right]\;. 
\label{nsmodepsi2}
\end{eqnarray}
The orientation $\theta$ is varied between $-45^0$ and $+45^0$ to provide a $0$ 
or $\pi$-class trajectory, respectively.  
A second $\pi/2$ mode converter (AMC-2) also oriented at $22.5^\circ$ generates 
\begin{eqnarray}
\mathbf{\Psi}_3&=& \left[ \mathbb{1}\otimes C(22.5^\circ,\pi/2)\right] \mathbf{\Psi}_2 \nonumber\\
&=& 
\frac{1}{\sqrt{2}}\left[
\begin{matrix}
-\sin2\theta &e^{-\frac{i\pi}{4}}&\\ 
-\cos2\theta &e^{\frac{i\pi}{4}}&\\ 
\;\;\;\cos2\theta &e^{-\frac{i\pi}{4}}&\\ 
-\sin2\theta &e^{\frac{i\pi}{4}}&
\end{matrix}
\right]\;, 
\label{nsmodepsi3}
\end{eqnarray}
and a quarter wave plate (QWP-2) oriented at $0^\circ$ makes the final transformation to 
\begin{eqnarray}
\mathbf{\Psi}_4 &=& \left[ W(0^\circ,\pi/2)\otimes\mathbb{1}\right] \mathbf{\Psi}_3 \nonumber\\
&=& 
\frac{1}{\sqrt{2}}\left[
\begin{matrix}
-&\sin2\theta& \\ 
-i&\cos2\theta& \\ 
-i&\cos2\theta& \\ 
-&\sin2\theta& 
\end{matrix}
\right]\;. 
\label{nsmodepsi4}
\end{eqnarray}
From (\ref{nsmodepsi4}) we immediately see that the complete transformation sequence 
is cyclic when $\theta=\pm 45^0\,$, giving $\mathbf{\Psi}_4=\pm\mathbf{\Psi}_0\,$. 

\begin{figure}
\begin{center} 
\includegraphics[scale=.5]{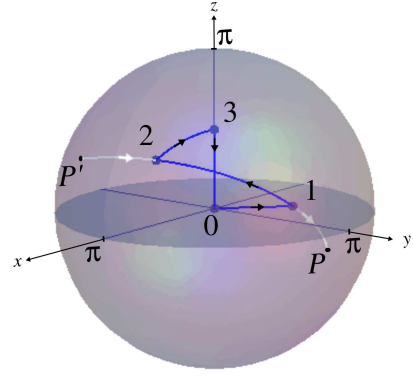}
\end{center} 
\caption{Diagram representing the transformations performed in the $SO(3)$ 
sphere. The $0-$class path is indicated in blue (online) and the $\pi-$class 
path between points 1 and 2 is indicated in white.} 
\label{spheres}
\end{figure}

In order to draw the trajectories in the SO($3$) sphere, we will use the parameterization 
detailed in ref.\cite{chines}. An arbitrary maximally non separable mode can be written as 
\begin{eqnarray}
\mathbf{\Psi}_{NS}&=&\frac{1}{\sqrt{2}}
\left(\lambda\,\psi_+\mathbf{\hat{e}}_H +
\eta\,\psi_-\mathbf{\hat{e}}_H -
\eta^*\,\psi_+\mathbf{\hat{e}}_V 
%\nonumber\\ &+&
+ \lambda^*\,\psi_-\mathbf{\hat{e}}_V\right)
\nonumber\\
&\equiv& \frac{1}{\sqrt{2}}\left[\lambda,\eta,-\eta^*,\lambda^*\right]^T\;. 
\label{gennsmode}
\end{eqnarray}
The point representing mode (\ref{gennsmode}) in the SO($3$) sphere is localized 
by the vector $\mathbf{v}=a\,\mathbf{\hat{u}}\,$, where $a\in[0,\pi]$ is the distance from the origin 
of the sphere and $\mathbf{\hat{u}}$ is the unit vector oriented from the origin to the 
point representing the corresponding mode. Following ref.\cite{chines}, the mode 
coefficients and the point coordinates are related by
\begin{eqnarray}
\lambda &=& \cos\frac{a}{2}-i\,u_z\sin\frac{a}{2}\;,
\nonumber\\
\eta &=& -(u_y+i\,u_x)\sin\frac{a}{2}\;.
\end{eqnarray}
Mode $\mathbf{\Psi}_0$ is represented by the origin of the SO($3$) sphere. 
Therefore, for $\theta=\pm 45^0\,$, the 
mode transformations are represented by the following points 
$(a,u_x,u_y,u_z)$ in the sphere
\begin{eqnarray}
\mathbf{\Psi}_0 &\equiv& \mathbf{\Psi}_4 \equiv (0,0,0,1)\;,
\nonumber\\
\mathbf{\Psi}_1 &\equiv& \left(\frac{\pi}{2},-\frac{1}{\sqrt{2}},\frac{1}{\sqrt{2}},0\right)\;,
\nonumber\\
\mathbf{\Psi}_2 &\equiv& \left(\frac{2\pi}{3},\sqrt{\frac{2}{3}},0,\frac{1}{\sqrt{3}}\right)\;,
\nonumber\\
\mathbf{\Psi}_3 &\equiv& \left(\frac{\pi}{2},0,0,1\right)\;.
\label{points}
\end{eqnarray}
These points are shown in Fig.(\ref{spheres}) together with the alternative 
paths connecting them. When HWP-2 is set to $\theta=-45^0\,$, the $0-$class 
trajectory indicated in blue (online) is followed from $\mathbf{\Psi}_1$ to 
$\mathbf{\Psi}_2\,$. When HWP-2 is set to $\theta=45^0\,$, the $\pi-$class 
trajectory indicated in white is followed, where the surface of the SO($3$) 
sphere is crossed at point $P$ and the trajectory continues from the opposite 
point $P^{\,\prime}\,$. The $\pi$ phase shift becomes 
clear when we compare Eqs.(\ref{nsmode}) and (\ref{nsmodepsi4}) for each 
value of $\theta\,$. In order to evidence this phase shift experimentally, 
we need to interfere the transformed beam with a reference. 

%
%============================================================================
%============================================================================
%

\begin{figure}
\begin{center} 
\includegraphics[scale=0.5]{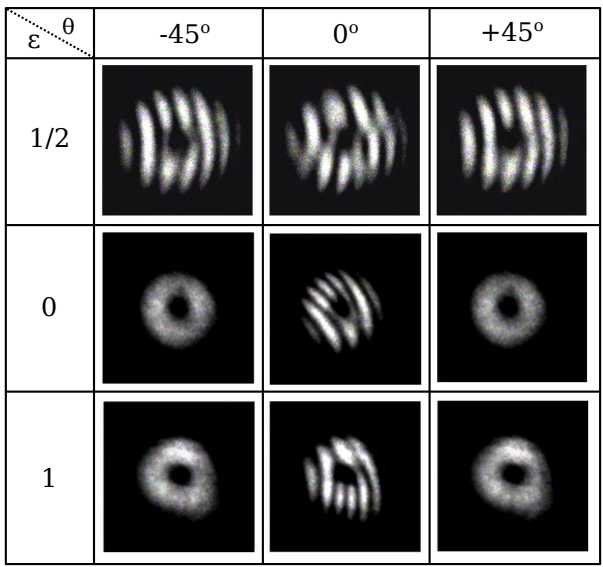}
\end{center} 
\caption{Images of the interference pattern observed at the output of the 
Mach-Zehnder interferometer. The corresponding values of $\epsilon$ and 
$\theta$ are indicated.}
\label{images-exp}
\end{figure}

The two arms of the 
interferometer are slightly misaligned to provide spatial interference. 
We can calculate the expected interference pattern assuming the general 
initial spin-orbit mode given by Eq.(\ref{initialnsmode}),
\begin{eqnarray}
I(\mathbf{r})&=&\left|\mathbf{\Psi}_4(\mathbf{r})+e^{i\,\mathbf{q}\cdot\mathbf{r}}\mathbf{\Psi}_{0}(\mathbf{r}) \right|^2\;,
\nonumber\\
&=& F(\rho,z)\,\left[1-2\sqrt{\epsilon\,(1-\epsilon)\,}\,\sin2\theta\,\cos(\mathbf{q}\cdot\mathbf{r})\right.
\nonumber\\
&-& \epsilon\,\cos2\theta\,\sin(\mathbf{q}\cdot\mathbf{r}+2\varphi)
\nonumber\\
&-& \left. (1-\epsilon)\,\cos2\theta\,\sin(\mathbf{q}\cdot\mathbf{r}-2\varphi)\right]\;, 
\label{interference}
\end{eqnarray}
where $\mathbf{q}$ is the transverse wavector difference between the 
misaligned beams and 
\begin{eqnarray}
F(\rho,z) &=& \frac{4\rho^2}{\pi\,w^4(z)}\,
\exp\left(-\frac{2\rho^2}{w^2(z)}\right)\;, 
\label{F}
\end{eqnarray}
is the radially symmetric intensity distribution of the Laguerre-Gaussian 
modes. Eq.(\ref{interference}) allows us to predict the effects of the 
parameters involved in the experimental setup. When HWP-2 is oriented at 
$\theta=\pm 45^0$ (the settings leading to closed trajectories in the 
SO($3$) sphere), the last two terms in the square brackets vanish and the 
interference pattern is determined solely by the second term with the 
fringes visibility given by 
\begin{equation}
\mathcal{V}(\theta=\pm 45^0) = \left|2\,\sqrt{\epsilon\,(1-\epsilon)\,}\right|\;. 
\label{visibility}
\end{equation}
Therefore, maximal visibility is expected for $\epsilon=1/2\,$ and 
no interference fringes are expected for separable modes 
($\epsilon=0,1$). This behavior was experimentally confirmed, as can be seen 
from the interference patterns shown in Fig.(\ref{images-exp}). The upper 
left and right corners of this figure correspond to the interference patterns 
obtained with the non separable mode for the $0$ and $\pi$ classes of closed trajectories. 
The $\pi$ phase 
shift is clear from the location of the phase singularity present in the LG modes, 
which is used as a spatial reference. For $\theta=-45^0\,$, the singularity is 
located on a bright fringe while for $\theta=45^0$ it falls on a dark fringe. 
The interferometer was stable over several minutes, more than enough for 
image acquisition and rotation of the wave plate from 
$-45^0$ to $45^0$ in steps of $22.5^0$, so that no active stabilization was 
required. The separable modes were obtained by selecting each polarization 
component with a polarizing beam splitter (PBS). As expected, no interference 
fringes are seen when $\theta=\pm45^0\,$, what can be verified in the left 
and right corners of the middle and bottom rows of Fig.(\ref{images-exp}). 

%
%============================================================================
%============================================================================
%

\begin{figure}
\begin{center} 
\includegraphics[scale=0.5]{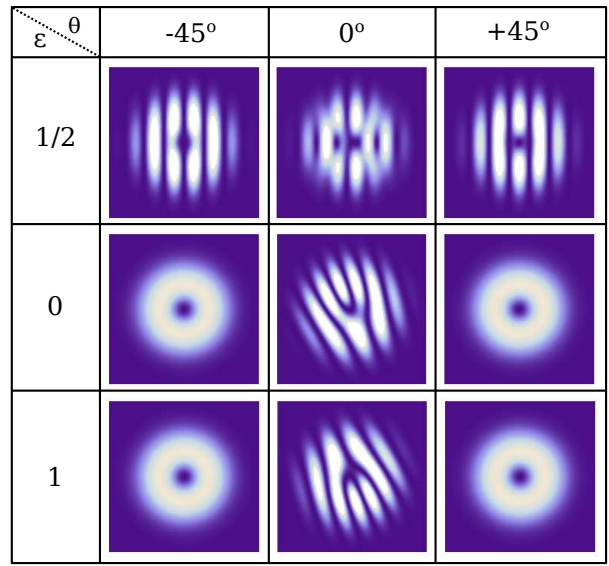}
\end{center} 
\caption{Density plot of Eq.(\ref{interference}), giving the theoretical images 
of the interference patterns at the output of the Mach-Zehnder interferometer. 
The images are in one to one correspondence with those in Fig.\ref{images-exp}.}
\label{images-theo}
\end{figure}

We could also simulate the whole set of interference patterns in a density 
plot of the theoretical expression (\ref{interference}). The corresponding 
results are shown in Fig.(\ref{images-theo}). The only fitting parameters are 
the components of the transverse wave vector $\mathbf{q}$, which were set to provide 
orientation and frequency of the interference fringes similar to those of the 
experimental patterns. The theoretical images exhibit a fairly good 
agreement with the experimental patterns in regards to the fringes visibility 
and their geometrical structure. The experimental patterns are slightly curved 
due to a small wave front mismatch in the Mach-Zehnder interferometer, 
not simulated in the theoretical images. The $\pi$ phase shift of the nonseparable 
mode is also evident from the displacement of the interference fringes with 
respect to the location of the phase singularity. 

%
%============================================================================
%============================================================================
%

\section{Conclusion}

In conclusion, we investigated the topological phase acquired by vector 
vortex beams under combined local unitary operations on the polarization 
and transverse mode degrees of freedom. The representation of maximally 
nonseparable spin-orbit modes in the SO($3$) manifold gives a crucial 
role to the doubly connected topology of the group in the 
interpretation of the possible phase values acquired under cyclic 
evolutions. Our measurements involve operations on both degrees of 
freedom in a Mach-Zehnder interferometer, and the role of the mode 
separability could be captured through the visibility of the 
interference patterns.

\acknowledgments
Funding was provided by Coordena\c c\~{a}o de Aperfei\c coamento de Pessoal 
de N\'\i vel Superior (CAPES), Funda\c c\~{a}o de Amparo \`{a} Pesquisa do 
Estado do Rio de Janeiro (FAPERJ-BR), and Instituto Nacional de Ci\^encia e 
Tecnologia de Informa\c c\~ao Qu\^antica (INCT-IQ-CNPq).

\end{document}